\documentclass[twocolumn,twoside,slac_two]{revtex4}
\usepackage{graphicx}
\usepackage{fancyhdr}
\newcommand{\order}{{\cal O}}

\newcommand{\nl}{\nonumber\\}

\newcommand{\be}{\begin{equation}}
\newcommand{\ee}{\end{equation}}

\newcommand{\bea}{\begin{eqnarray}}
\newcommand{\eea}{\end{eqnarray}}

\begin{document}

\title{
Update on Semileptonic Charm Decays
}

%

\author{Richard J. Hill}
\affiliation{
Fermi National Accelerator Laboratory\\
\it P.O. Box 500, Batavia, Illinois 60510, USA
}

\begin{abstract}
A brief update is given on recent developments in
the theory of exclusive semileptonic charm decays.  
A check on analyticity arguments from the kaon system is
reviewed.  
Recent results on form factor shape measurements are discussed. 
\end{abstract}

\maketitle

\thispagestyle{fancy}


\section{Introduction}

Semileptonic meson decays provide a valuable arena 
to study the weak and strong interactions.   
On the one hand, once the effects of the strong interaction are under control, 
weak mixing parameters ($V_{ub}$, $V_{us}$, ... )  
can be extracted from the overall normalization.   On the other hand, 
if these mixing 
parameters are taken from other processes, then the semileptonic 
rates probe complicated underlying strong dynamics, thus yielding an  
important test for lattice methods~\cite{aida} and for our understanding of 
nonperturbative QCD.

Apart from the overall normalization, the spectral shapes are also 
interesting~\cite{Hill:2006ub}.
The energy spectra (the ``$q^2$ dial'') are governed by 
quark-hadron duality, and are constrained by dispersion relations and
analyticity.   
Shape parameters for different quark masses (the ``$m$ dial'')
can be interpreted by means of effective field theory descriptions in the 
appropriate regimes of validity.  
These parameters provide inputs to 
other processes governed by the same effective field theory.   
For example, the form factors measured in 
$B\to\pi\ell\nu$ constrain $B\to\pi\pi$.  

The charm quark holds a privileged position in this scheme.  
Abundant experimental data exist for $D\to K$ and for $D\to \pi$ 
transitions.   The charm mass is heavy enough to be treated using
heavy-quark methods on the lattice, but light enough so that the full
range of physically allowed momentum transfers are accessible in present
simulations~\cite{Aubin:2004ej}.   
Charm decays thus provide a powerful test of lattice QCD methods that 
can be applied to other heavy-meson systems.

Charm decays are important for testing another important, but
perhaps less well-known aspect of QCD, namely 
the constraints imposed by 
analyticity~\cite{Bourrely:1980gp,Fukunaga:2004zz,Arnesen:2005ez,
Becher:2005bg,Flynn:2006vr}.   
These constraints
imply a convergent expansion in powers of a small parameter that 
measures the distance between the physically allowed kinematic region, 
and the region of resonances and production thresholds.  
While the existence of this small parameter is well known, the usefulness of the expansion
has been practically negated by appeals to ``unitarity bounds''.
It has only recently become clear with the advent of rigorous
power-counting arguments that we can ``take seriously'' the expansion
provided by analyticity~\cite{Becher:2005bg}.   
Charm decays provide a valuable illustration and crosscheck of 
these arguments.   

An overview of these ideas for general semileptonic meson transitions 
has been presented previously in \cite{Hill:2006ub}; 
further discussion and references may be found there. 
The focus in the present report is on some recent illustrations 
from kaon physics that are relevant to charm physics, 
and on an update of shape measurements in the charm system.  

The remainder of the talk is organized as follows.  Section~\ref{sec:kaon} 
outlines the constraints of analyticity, and describes the explicit test 
of power-counting afforded by $K_{\ell 3}$ decays.   
Section~\ref{sec:results} tabulates recent results on the shape parameters 
in $D\to K$ and $D\to \pi$ semileptonic transitions.   
Section~\ref{sec:conclude} concludes with a discussion of the relevance of 
these shape parameters for testing lattice QCD, and for 
applying factorization in $B$ decays.

\section{Analyticity and simplicity\label{sec:kaon}}

\begin{table}[t]
\begin{center}
\caption{
Maximum $|z|$ throughout semileptonic range (from \cite{Hill:2006ub}). 
}
\begin{tabular}{ccc}
\textbf{\quad Process \quad}  & \textbf{CKM element} & \textbf{$|z|_{\rm max}$}
\\
\hline 
$\pi^+ \to \pi^{0}$ & $V_{ud}$ & $3.5 \times 10^{-5}$ \\
$B \to D$ & $V_{cb}$ & 0.032 \\
$K \to \pi$ & $V_{us}$ & 0.047 \\
$D \to K$ & $V_{cs}$ & 0.051 \\
$D \to \pi$ & $V_{cd}$ & 0.17 \\
$B \to \pi$ & $V_{ub}$ & 0.28 
\end{tabular}
\label{tab:zmax}
\end{center}
\end{table}

The mere existence of a field theory description of the physical 
hadrons, and their weak current probes, implies powerful constraints on
form factors.  In particular, singularities in hadronic transition
amplitudes are determined by kinematics.  Analyticity translates
into a convergence expansion in a small variable once the domain 
of analyticity is mapped onto a standard region.   

In practical terms, this mapping is simply a rearrangement of the
series expansion of the form factor,
\begin{equation}
F(t) = F(0) + \dots \,. 
\end{equation}
For example, a simple pole model of the form factor would ``resum''
into the form
\begin{equation}
F(t)/F(0)= 1 + t/m^2 + (t/m^2)^2 + \dots = \frac{1}{1-{q^2/m^2}} \,. 
\end{equation}
Without knowing what the form factor ``resums'' into, analyticity 
implies that the series has to rearrange itself into the form
\begin{equation}
\label{eq:expand}
F(t)/F(0) = 1 + a_1 z(t) + a_2 z^2(t) + \dots \,, 
\end{equation} 
where $a_i$ are $\order(1)$ in a rigorous sense 
($\sum_i a_i^2$ is also $\order(1)$), and $z$ is a variable
bounded by the distance of the physical region from singularities.

The smallness of $|z|$ in (\ref{eq:expand}) 
implies that terms beyond linear order are highly suppressed. 
There is thus an essentially unique choice
of shape parameter that unites semileptonic transitions from various
decay modes---the slope of the form factor, say at $q^2=0$.  
It turns out that the numerical value of this parameter is in itself an
interesting quantity.   It provides a quantitative test of lattice
versus experimental shape; it is an input to related hadronic processes; 
and it probes an unsolved question in the QCD dynamics governing form
factors~\cite{Hill:2006ub}.

\begin{figure}
\includegraphics[width=80mm]{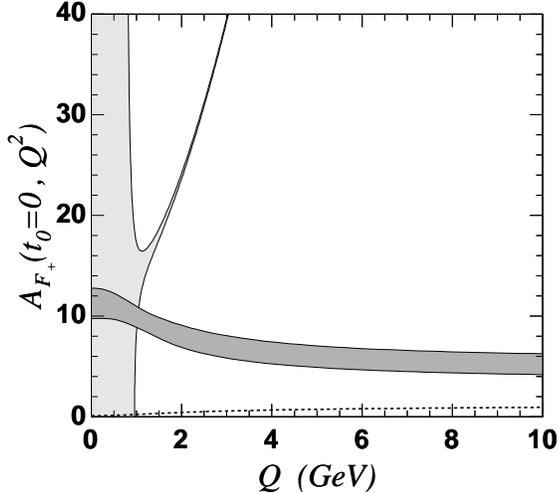}
\caption{\label{fig:plotT} Sum of squares of expansion coefficients (dark) compared to
unitarity bound (light).  (from \cite{Hill:2006bq}) } 
\end{figure}

The tools of analyticity are well-known, but their usefulness
has not been appreciated, due to a reliance on 
unitarity as the only means to bound the coefficients $a_k$.  
In fact, rigorous power counting arguments show that both $a_k$ and 
$\sum_k a_k^2$ are order unity, even when large ratios of scales, such as 
$m_Q/\Lambda_{\rm QCD}$, are present 
($m_Q$ a heavy-quark mass, and $\Lambda$ a hadronic scale). 
It is important to test and make full use of this expansion when dealing with
complicated QCD systems. 

Analysis of $K_{\ell 3}$ decays ($K\to\pi\ell\nu$) provides a revealing
confirmation of these ideas.  The analysis also provides an important constraint on the
determination of $|V_{us}|$~\cite{Hill:2006bq}.   
Figure~\ref{fig:plotT} compares the unitarity 
bound to the actual value of the sum of squares of coefficients, as 
calculated from ALEPH $\tau$ decay data~\cite{Barate:1999hj}.  
The result is plotted 
for a range of values of the OPE parameter $Q$ ($Q$ corresponds to the 
virtuality of the produce meson pair in the crossed channel; large $Q$ implies
that the OPE is increasingly reliable). 
It can be seen that the unitarity bound begins to wildly overestimate the 
possible size of the coefficients for even moderately large $Q$.   
The convergence of the series improves as $Q$ becomes larger,
since the sensitivity to the $K^*$ pole is lessened; 
reliance on the unitarity bound would however force 
us to work at small $Q$.  

\section{\label{sec:results}Shape parameters}  

\begin{figure}[h]
\centering
\includegraphics[width=80mm]{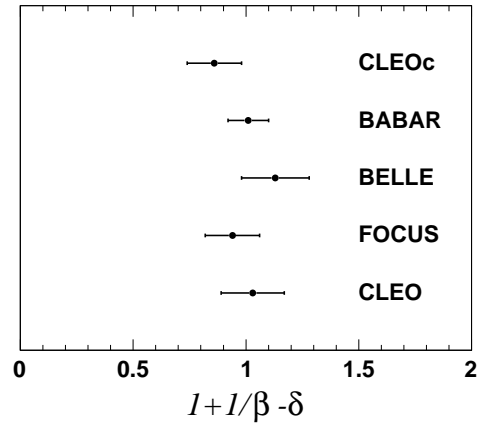}
\caption{ Relative slope of the vector form factor in $D\to K\ell\nu$. 
As tabulated in (\ref{eq:DK}) and \cite{Hill:2006ub}, from 
CLEO~\cite{Huang:2004fr},
FOCUS~\cite{Link:2004dh},
BELLE~\cite{Abe:2005sh},
BABAR~\cite{Aubert:2007wg}
and CLEOc~\cite{:2007se}. 
\label{fig:DK} }
\end{figure}

\begin{figure}[h]
\centering
\includegraphics[width=80mm]{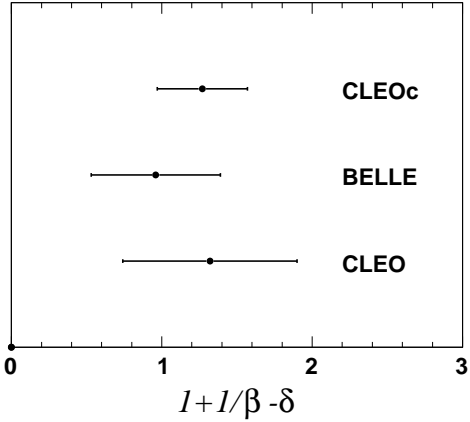}
\caption{ Relative slope of the vector form factor in $D\to \pi\ell\nu$. 
As tabulated in (\ref{eq:Dpi}) and \cite{Hill:2006ub}, from 
CLEO~\cite{Huang:2004fr},
BELLE~\cite{Abe:2005sh},
and CLEOc~\cite{:2007se}
\label{fig:Dpi} }
\end{figure}

Given the rapid convergence of the series (\ref{eq:expand}), it
turns out that no more than a normalization and slope can be measured
in present experiments~\cite{Hill:2006ub}.   
It is convenient to define the slope at $q^2 = 0$, in which case
we have: 
\begin{eqnarray}\label{eq:betadef} 
{1\over \beta} &\equiv& 
{m_H^2-m_L^2 \over F_+(0)} \left. dF_{0} \over dt\right|_{t=0} \,, \nl
\delta &\equiv& 
1 - {m_H^2-m_L^2 \over F_+(0)} 
\left( \left. dF_{+} \over dt\right|_{t=0}  
- \left. dF_{0} \over dt\right|_{t=0} \right) \nl
&\equiv& {F_+(0)+F_-(0)\over F_+(0)}
\,.
\end{eqnarray}
Here $F_+$ is the vector form factor that dominates for massless leptons in
pseudoscalar-pseudoscalar transitions, and $F_0$ is the scalar form factor, 
accessible for massive leptons, or in lattice simulations. 

In the heavy-quark limit of heavy-to-light form factors, 
soft-collinear effective theory~\cite{Bauer:2000ew} can be used to extract the 
heavy-quark mass dependence, and organize form-factor contributions in powers
of $\Lambda_{\rm QCD}/E$, where $E$ is the energy of the outgoing light hadron. 
At leading power~\cite{Hill:2005ju}, 
\begin{eqnarray}\label{eq:scetff}
F_+(E) &=& \sqrt{m_H}\big[ \zeta(E) + \left({4E\over m_H} - 1 \right) H(E) \big] \,, \nl
{m_H\over 2E} F_0(E) &=& \sqrt{m_H}\big[ \zeta(E) + H(E) \big] \,. 
\end{eqnarray}
Here $\zeta$ is the so-called soft overlap contribution to the form factor, 
and $H$ is the hard-scattering component, due to hard gluon exchange with the 
spectator quark. 
Apart from scaling violations, the functions $\zeta$ and 
$H$ both have an energy dependence $\zeta\sim H\sim E^{-2}$. 
Then independent of any model assumptions, 
\be
{1\over\beta} = -\left. {d \ln(\zeta + H ) \over d\ln E} \right|_{E=m_H/2} - 1 
+ \order( \alpha_s, \Lambda/m_H ) \,,  
\ee
and 
\begin{eqnarray}\label{eq:deltascet}
\delta(m_L,m_H) &=& \left. { 2 H \over \zeta + H } \right|_{E=m_H/2} 
+ \order( \alpha_s, \Lambda/m_H ) \,. 
\end{eqnarray} 
Thus $1/\beta - 1$ measures scaling violations, and $\delta$ measures the size
of hard-scattering contributions. 

From the recent BABAR~\cite{Aubert:2007wg} and CLEOc~\cite{:2007se} 
results on $D\to K$, 
\bea\label{eq:DK}
\boldmath{ D\to K }: \quad\quad\quad && \nl 
1+1/\beta-\delta &=& 1.01 \pm 0.04 \pm 0.08 \quad \text{\cite{Aubert:2007wg}} \nl
&& 0.85 \pm 0.06 \pm 0.10 \quad \text{\cite{:2007se}} \,. 
\eea
To compare with \cite{Hill:2006ub}, the central values and first error 
correspond to keeping just the linear term ($a_0$ and $a_1$)
in (\ref{eq:expand}), and the second error is a conservative 
bound on residual shape uncertainty from allowing $a_2$ with
$|a_2/a_0| \le 10$. 
Together with previous CLEO, FOCUS and BELLE measurements, these values 
are displayed in Figure~\ref{fig:DK}. 
A naive average of these results yields
$1+1/\beta - \delta = 0.97 \pm 0.05$  ($\chi^2 = 2.5$ for 5-1 d.o.f.).  

From the CLEOc results on $D\to \pi$, 
\bea\label{eq:Dpi}
\boldmath{ D\to \pi }: \quad\quad\quad && \nl 
1+1/\beta-\delta &=& 
1.27 \pm 0.13 \pm 0.27 \quad \text{\cite{:2007se}} \,. 
\eea
Together with previous 
CLEO and BELLE measurements, these values 
are displayed in Figure~\ref{fig:Dpi}.  
A naive average of these results yields
$1+1/\beta - \delta = 1.19 \pm 0.23$  ($\chi^2 = 0.4$ for 3-1 d.o.f.).  

\section{\label{sec:conclude} Conclusions} 

Semileptonic charm decays provide an important arena in which to test 
lattice QCD.  
Unfortunately, it is not possible to definitively test the experimental 
shape predictions against unquenched lattice simulations at present, since
the lattice results are so far reported only in terms of model parameters, 
that need not agree between theory and experiment, or between
experiments with different acceptance and systematics.  It is interesting
to note that while some model parameters show poor agreement between 
experiments~\cite{Wiss:2006ih}, there is no obvious discrepancies among the
physical results shown in Figure~\ref{fig:DK}. 

The actual value of the shape parameters are also of interest.   
For example, the parameter $\delta$ is an important input to factorization
analyses of $B\to\pi\pi$ decays~\cite{Bauer:1986bm}, 
usually phrased in terms of an (inverse) moment
of the $B$ meson wavefunction, $\lambda_B$: 
\be
\delta(m_\pi,m_B) = {6\pi C_F \over N_c} 
{f_B f_\pi \alpha_s \over m_B \lambda_B F_+(0)} + \dots \,.
\ee
A plausible conjecture of monotonicity for $\delta$ implies 
that $\delta(m_\pi,m_B) < 0.35 \pm 0.03$.
A much larger value of $\delta(m_\pi,m_B)$ in $B\to \pi$, 
e.g. $\delta \approx 1$~\cite{Bauer:2004tj}, 
would require a dramatic behavior
of this physical quantity as a function of quark mass, a behavior that should
already be apparent in the charm system.   For example, 
we would expect $\delta(m_\pi, m_D) > \delta(m_K, m_D)$ (see Figure~8 of 
\cite{Hill:2006ub}). 
It is exciting that current lattice simulations can probe this quantity 
in both the charm and bottom systems, not by evaluating moments of $B$ meson
wavefunctions, but by computing form factor slopes.   Experimental measurements
are sensitive to the combination $1/\beta - \delta$, so that the scalar form
factor slope (denoted $1/\beta$) is the most urgent requirement from the lattice. 
Again, definitive conclusions must await a model-independent presentation of 
simulation results. 

\begin{acknowledgments}
Research supported by the U.S.~Department of Energy  
grant DE-AC02-76CHO3000.
\end{acknowledgments}

\bigskip 

\end{document}